# Chaotic Quantum Transport near the Charge Neutrality Point in Inverted Type-II InAs/GaSb Field-Effect Transistors

W. Pan, J.F. Klem, J.K. Kim, M. Thalakulam, M.J. Cich

Sandia National Laboratories, Albuquerque, New Mexico 87185, USA

(March 27, 2012)

Abstract:

We present here our recent quantum transport results around the charge neutrality point (CNP) in a type-II InAs/GaSb field-effect transistor. At zero magnetic field, a conductance minimum close to $4e^2/h$ develops at the CNP and it follows semi-logarithmic temperature dependence. In quantized magnetic (B) fields and at low temperatures, well developed integer quantum Hall states are observed in the electron as well as hole regimes. Quantum transport shows noisy-like behavior around the CNP at extremely high B fields. Surprisingly, when the diagonal conductivity $\sigma_{xx}$ is plotted against the Hall conductivity $\sigma_{xy}$, a circular conductivity law is discovered, suggesting a chaotic quantum transport behavior.

Research on topological insulator (TI) has drawn intensive interests [1,2]. This material system is believed to hold great potential for future quantum computation [3,4]. Among the many known TI systems, the type-II InAs/GaSb heterostructure appears to be most technology mature in terms of material growth. Indeed, a large amount of work has already been carried out in this material system for infrared applications [5]. It is well known that in this type II heterostructure there exists an unusual alignment of the conduction and valence band edges of InAs and GaSb [6], with the valence band edge of GaSb higher than the conduction band edge of InAs (shown schematically in Figure 1a). Due to this band inversion, a hybridization gap forms (as shown in Fig. 1b) and, at some critical quantum well thickness, the system was theoretically predicted to become a two-dimensional (2D) TI. It can support dissipationless time-reversal symmetry protected edge channels and display the quantum spin Hall effect (QSHE) [7,8]. In less than 4 years after this theoretical prediction, there has been experimental evidence [9] supporting the existence of QSHE.

Very early on, it was realized that the electronic transport at the charge neutrality point (CNP), where electrons and holes coexist, is of quantum nature [10]. Recently, novel spin states inside the hybridization gap have been predicted [11] for spintronic device applications. Under non-zero magnetic (B) fields, non-trivial magneto-conductance was also studied theoretically in the low field regime, where the cyclotron energy is smaller than the Fermi energy [10]. Little is known, however, experimentally and theoretically on transport behaviors in the extreme quantum limit, where the cyclotron energy is much larger than the Fermi energy. On the other hand, due to strong carrier-carrier interactions and correlations, novel physics can be expected at very high B fields in the CNP regime.

In this Letter, we present our recent quantum transport results around the CNP in a type-II InAs/GaSb field-effect transistor. At zero magnetic (B) field, a conductance minimum close to $4e^2/h$ develops at the CNP and it follows semi-logarithmic temperature dependence. Under high B fields, well developed integer quantum Hall states, manifested by precisely quantized Hall plateaus and vanishingly small diagonal magnetoresistance, are observed in the electron as well as hole regimes. Quantum transport shows noisy-like behavior around the CNP at ultra-high B fields. Surprisingly, when the diagonal conductivity $\sigma_{xx}$ is plotted against the Hall conductivity $\sigma_{xy}$, a peculiar circular conductivity law is discovered, suggesting a chaotic quantum transport behavior.

We basically followed the recipe in Ref. [12] for the growth structure of our InAs/GaSb heterostructures, except that the well width of InAs is changed to 150Å. Field-effect transistors (FETs) were fabricated using conventional photolithography techniques [13]. A microscopic image of one InAs/GaSb FET is shown in Fig. 1c. The Hall bar width (W) is 10 μm, and the distance between two voltage leads (L) is 30 μm. Magneto-transport measurements were carried out either in a $^3$He/$^4$He dilution refrigerator equipped with a superconducting magnet at Sandia National Laboratories or a portable dilution refrigerator with a resistive magnet at the National High Magnetic Field Laboratory (in Tallahassee, Florida, USA) when in need of ultra-high B fields (>20T). Conventional low frequency (~ 11 Hz) phase lock-in techniques were used to measure the transport coefficients, the diagonal resistance $R_{xx}$ and Hall resistance $R_{xy}$. The diagonal resistivity $\rho_{xx}$ is obtained after taking into account the geometric ratio, $\rho_{xx} = W/L \times R_{xx}$. $\rho_{xy} = R_{xy}$ in the two-dimensional carrier systems. The conductivity tensor components are then determined by $\sigma_{xx} = \rho_{xx}/(\rho_{xx}^2 + \rho_{xy}^2)$ and $\sigma_{xy} = \rho_{xy}/(\rho_{xx}^2 + \rho_{xy}^2)$. Two devices from the same wafer were examined and the results from these two specimens are consistent with each other.

Fig.1d shows $R_{xx}$ and $R_{xy}$ as a function of front gate voltage $V_g$ at zero B field. $R_{xx}$ first increases slowly as $V_g$ is decreased from ~ 9V. Around $V_g \sim 2V$, it increases quickly and develops a maximum at $V_g \sim -2.5V$. Then $R_{xx}$ decreases with $V_g$ further decreasing. This $V_g$ dependence of $R_{xx}$ is consistent with previous work [9, 14] and confirms the existence of two-carrier transport in our device. $R_{xy}$ is basically zero at large positive and negative $V_g$. The non-zero value around the CNP of $V_g \sim -2.5V$ (as we will discuss in detail in the following) and the reversal of its sign are similar to those in graphene around the Dirac point [15], indicating switching of one type of carriers to another.

In Figure 2a, we show the conductance, obtained by $G_{xx} = 1/R_{xx}$, around $V_g \sim -2.5V$. A minimum very close of $4e^2/h$ was observed at $V_g = -2.45V$. This gate voltage is very close to the charge neutrality point (CNP) at $V_g = -2.47V$, as we will discuss in detail as follows. Temperature (T) dependence of the minimal conductance was carried out. Same as in previous work [9,14], the temperature dependence is much weaker than the exponential dependence expected for a system with a hard energy gap. In fact, the conductance follows a semi-logarithmic dependence and in this sample $G_{xx} = 3.97 + 0.10 \times \log(T)$, where $G_{xx}$ is in units of $e^2/h$ and T in mK (Fig. 2b).

Under quantized B fields, the integer quantum Hall effect (IQHE) is observed. Figure 3a shows $\sigma_{xx}$ and $\sigma_{xy}$ versus $V_g$ at B = 5T. Quantized Hall plateaus can be seen at Landau level fillings $\nu_e$=4, 5, 6, ... to 12 in the electron regime. Corresponding to these Hall plateaus, vanishingly small $R_{xx}$ is observed at even fillings and well defined $R_{xx}$ minima at odd fillings. Furthermore, developing quantum Hall states are seen at $\nu_e$=1 and 2 in the electron side and $\nu_h$=-2 and -4 in the hole side, evidenced by the kinks in the Hall resistance and minima in $R_{xx}$. It is also worthwhile pointing out that $R_{xy}$ assumes an almost zero value over a finite range of $V_g$ around the CNP and, concomitantly, $R_{xx}$ displays a minimum. We note that similar transport features were observed in a CdHgTe/HgTe/CdHgTe heterostructure [16], also a two-dimensional TI, at a magnetic field of similar strength. There, it was proposed that the transport at the CNP was dominated by electron-hole "snake states" propagating along the $\nu = \nu_e + \nu_h = 0$ lines.

Electron (hole) density can be deduced from the positions of $R_{xx}$ minima in the quantum Hall effect according to $n(p) = \nu_e(\nu_h)eB/h$, and plotted as a function of $V_g$ in Fig. 3b. It is clearly seen that, there, the positions of the zero electron (n) and hole (p) density do not overlap at the same $V_g$. In fact, n = 0 occurs at $V_g$=-2.91V and p = 0 at $V_g$=-2.02V from the extrapolations (red straight lines). As a result, at the CNP (or n+p=0), the electron and hole densities are not exactly zero but finite and $|n| = |p| \sim 0.6 \times 10^{11}$ cm$^{-2}$.

In order to understand the physics origin of the minimal conductance, we need to determine the value of the hybridization gap $\Delta$. For this purpose, we carried out transconductance measurement [17]. Fig. 3c shows the result. The peak at $V_g^c$ = -1.45V corresponds to the position of the bottom of the hybrid "conduction" band, while the dip at $V_g^v$ = -2.69 V corresponds to the top of the hybrid "valance" band. The hybridization gap $\Delta$ of ~ 1 meV can be deduced using the formula $\Delta = (V_g^c - V_g^v) \times \Delta n/\Delta V \times (m_e+m_h)/\pi\hbar^2$ [9]. Here, $\Delta n/\Delta V$ is taken to be $1.33 \times 10^{11}$ cm$^{-2}$/V, the average of the rates of electron density change and hole density change with $V_g$, as shown in Fig. 3b. The electron and hole effective masses are $m_e$ = 0.03 $m_0$ and $m_h$ = 0.37 $m_0$, respectively, where $m_0$ is the bare electron mass. The value of $\Delta$ ~ 1 meV is about a quarter of those obtained in previous work [9,12].

With the above vale of $\Delta$, we determined that the formation of the minimal conductance (or conductivity) in Fig. 2a is most likely due to a bulk transport at the CNP. Indeed, as shown in Ref. [10], the bulk conductivity is given by $\sigma^{th}_{xx} \approx e^2/h \times E_{g0}/\Delta$, where $E_{g0}$, ~ 15 meV in our device, is the energy difference between the first hole and electron levels, as shown in Fig. 1b. Using the value of $\Delta$ ~ 1 meV, $\sigma^{th}_{xx}$ ~ 15 $e^2$/h, which is very close to the measured one $\sigma_{xx}$ = L/W $\times$ $G_{xx}$ = 12 $e^2$/h. This bulk transport in macroscopic devices has also been reported by Knez et al [9].

Alternatively, we note that a minimal conductivity at the CNP has also been observed in graphene layers [15]. There, it is generally believed that this is due to the formation of electron and hole puddles at the CNP [18]. As shown in Fig. 3b, in our device the electron and hole density is finite at the CNP. It is possible that these electrons and holes form intervened electron/hole puddles, which then gives rise to a finite minimal conductivity at the CNP.

Finally, we note that the weak temperature dependence of $G_{xx}$ at the CNP is consistent with a very small hybridization gap, ~ 1meV. The energy level broadening may further reduce the hybridization gap and makes the system similar to a semimetal, where a semi-logarithmic temperature dependence of $G_{xx}$ is well known in the diffusive transport regime [19].

We now focus on the magneto-transport near the CNP in the extreme quantum limit. The magneto-transport coefficients (diagonal resistivity $\rho_{xx}$ and Hall resistivity $\rho_{xy}$) at the magnetic (B) field of 20 Tesla's are shown in Fig. 4a and the diagonal conductivity $\sigma_{xx}$ and Hall conductivity $\sigma_{xy}$ in Fig.4b. Well developed IQHE states are seen at the Landau level filings $\nu_e$ = 1, 2, 3 in the electron regime and at $\nu_h$ = -1 and -2 in the hole regime. We wish to point out that the $\nu_h$ = -1 and -2 IQHE states are observed for the first time in type-II InAs/GaSb heterostructures, attesting the high quality of our sample.

In this extreme quantum limit both $\rho_{xx}$ ($\sigma_{xx}$) and $\rho_{xy}$ ($\sigma_{xy}$) display noisy transport behavior in the CNP regime. This is dramatically different from smaller magnetic fields, for example at B = 5T in Fig. 3a, where $\sigma_{xx}$ and $\sigma_{xy}$ vary smoothly with $V_g$ and no noisy behavior is observed. Furthermore, unlike the correlated fluctuation observed in mesoscopic samples in the quantum Hall transition regime [20], here, no correlation between $\rho_{xx}$ and $\rho_{xy}$ (or between $\sigma_{xx}$ and $\sigma_{xy}$) can be identified. Instead, a totally new, unexpected circular conductivity law is discovered. Figure 5 displays our surprising result in this Letter. When plotting $\sigma_{xx}$ as a function of $\sigma_{xy}$, it is clearly seen that the majority data points center around the circles defined by $(\sigma_{xx} - N)^2 + \sigma_{xy}^2 = N^2$, where $\sigma_{xx}$, $\sigma_{xy}$, and N are in units of $e^2/h$. This circular conductivity law only holds at high magnetic fields in the extreme quantum limit and becomes distorted at lower B fields.

The circular conductivity law has to be due to the co-existence of both electrons and holes and the interactions between these two types of carriers. Indeed, at the CNP both the electron and hole densities are finite and small, ~ $0.6 \times 10^{11}$ cm$^{-2}$. As a result, the Landau level filling factor for the electrons, as well as for the holes, is less than 1/5 at B = 20T and higher. Without strong interactions between the electrons and holes, both the 2D electron system and the hole system are expected to become localized [21,22]. Consequently, $\sigma_{xx}$ and $\sigma_{xy}$ are zero and no circular law is expected.

The fall of the data points onto discrete circles strongly resembles the chaotic transport in a wide quantum well under tilt magnetic field, where the breakup of the stable orbits gives rise to chaotic motions [23]. This mechanism may explain the chaotic transport in our device around the CNP, where electrons and holes coexist. In high magnetic fields, the edge channels of electrons and holes can be mixed up by the sample disorder and form eddy-like circles [24]. It is possible that the breakup of perfect, dissipationless edge states into discrete eddies is responsible for the chaotic transport behaviors. Why this chaotic transport results in an apparent circle conductivity law still needs to be understood. As an alternative explanation, we note that quantum chaotic behavior has also been predicted in disordered graphene [25,26]. It was theoretically predicted that, in nanoflakes with weak diagonal disorder induced by random electrostatic potential [26], there exists a transition to quantum chaos with increasing disorder strength as the statistical distribution of energy levels evolves from Poissonian to Wigner. More theoretical modeling and experimental examination are needed to verify whether this physical origin is relevant to the chaotic transport behaviors observed in our InAs/GaSb system.

In summary, we have discovered an exotic quantum chaotic transport behavior and a new circular conductivity law around the charge neutrality point in the extreme quantum limit in a type-II InAs/GaSb field-effect transistor. A better understanding of this new transport behavior is expected to lead to novel electron physics in topological insulators.

This work was supported by the DOE Office of Basic Energy Sciences. We thank J.A. Simmons and R.R. Du for very helpful discussions. Ultra-high magnetic field measurements were performed at the National High Magnetic Field Laboratory, which is supported by NSF Cooperative Agreement under Grant No. DMR-0084173, by the State of Florida, and by the DOE. Sandia National Laboratories is a multi-program laboratory managed and operated by Sandia Corporation, a wholly owned subsidiary of Lockheed Martin Corporation, for the U.S. Department of Energy's National Nuclear Security Administration under contract DE-AC04-94AL85000.

References:
[1] M.Z. Hasan and C.L. Kane, Rev. Mod. Phys. **82**, 3045 (2010).
[2] X.L. Qi and S.C. Zhang, Rev. Mod. Phys. **83**, 1057 (2011).
[3] L. Fu and C.L. Kane, Phys. Rev. Lett. **100**, 096407 (2008).
[4] C. Nayak, S.H. Simon, A. Stern, M. Freedman, and S. Das Sarma, Rev. Mod. Phys. **80**, 1083 (2008).
[5] M.M.Z. Tidrow, Infrared Phys. &Tech. **52**, 322 (2009).
[6] M. Altarelli, Phys. Rev. B **28**, 842 (1983).
[7] C.X. Liu, T.L. Hughes, X.L. Qi, K. Wang, and S.C. Zhang, Phys. Rev. Lett. **100**, 236601 (2008).
[8] M. König, S. Wiedmann, C. Brüne, A. Roth, H. Buhmann, L.W. Molenkamp, X.L. Qi, and S.C. Zhang, Science **318**, 766 (2007).
[9] I. Knez, R.R. Du, and G. Sullivan, Phys. Rev. Lett. **107**, 136603 (2011).
[10] Y. Naveh and B. Laikhtman, Europhys. Lett. **55**, 545 (2001).
[11] J. Li, W. Yang, and K. Chang, Phys. Rev. B **80**, 035303 (2009).
[12] M.J. Yang, C.H. Yang, B.R. Bennett, and B.V. Shanabrook, Phys. Rev. Lett. **78**, 4613 (1997).
[13] M.J. Yang, F.C. Wang, C.H. Yang, B.R. Bennett, and T.Q. Do, Appl. Phys. Lett. **69**, 85 (1996).
[14] L.J. Cooper, N. Patel, V. Drouot, E. Linfield, D. Ritchie, and M. Pepper, Phys. Rev. B **57**, 11915 (1998).
[15] K. S. Novoselov, A. K. Geim, S. V. Morozov, D. Jiang, Y. Zhang, S. V. Dubonos, I.V. Grigorieva, and A. A. Firsov, Science **306**, 666 (2004).
[16] G.M. Gusev, E.B. Olshanetsky, Z.D. Kvon, N.N. Mikhailov, S.A. Dvoretsky, and J.C. Portal, Phys. Rev. Lett. **104**, 166401 (2010).
[17] F.F. Fang and A.B. Fowler, Phys. Rev. **169**, 619 ((1968).
[18] S. Adam et al., PNAS **104**, 18392 (2007).
[19] B.L. Altshuler, A.G. Aronov, and P.A. Lee, Phys. Rev. Lett. 44, 1288 (1980).
[20] E. Peled, D. Shahar, Y. Chen, E. Diez, D.L. Sivco, and A.Y. Cho, Phys. Rev. Lett. **91**, 236802 (2003).
[21] H.W. Jiang, R.L. Willett, H.L. Stormer, D.C. Tsui, L.N. Pfeiffer, and K.W. West, Phys. Rev. Lett. **65**, 633 (1990).
[22] M.B. Santos, Y.W. Suen, M. Shayegan, Y.P. Li, L.W. Engel, and D.C. Tsui, Phys. Rev. Lett. **68**, 1188 (1992).
[23] G. Müller, G.S. Boebinger, H. Mathur, L.N. Pfeiffer, and K.W. West, Phys. Rev. Lett. **75**, 2875 (1995).
[24] R.J. Nicolas, K. Takashina, M. Lakrimi, B. Kardynal, S. Khym, N.J. Mason, D.M. Symons, D.K. Maude, and J.C. Portal, Phys. Rev. Lett. **85**, 2364 (2000).
[25] I. Amanatidis and S.N. Evangelou, Phys. Rev. B **79**, 205420 (2009).
[26] A. Rycerz, arXiv:1112.5078.

Figures and figure captions:

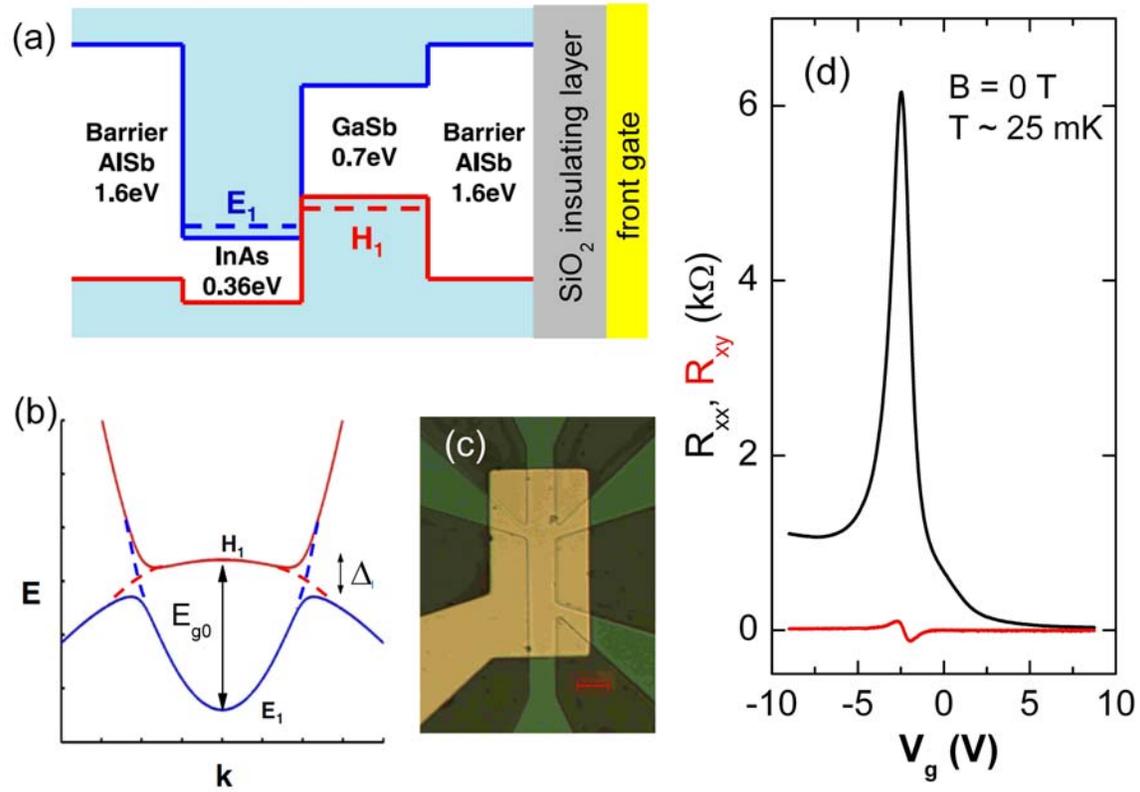

Figure 1 (color online) (a) Schematic band diagram of type-II InAs/GaSb heterostructure. (b) Schematic energy diagram in the inverted regime. Fig. 1(a) and (b) are adapted from Ref. [7]. (c) Optical micrograph of a type-II InAs/GaSb field-effect transistor. The scale bar is 10 μm. (d) $R_{xx}$ (black) and $R_{xy}$ (red) as a function of the front gate voltage $V_g$ at B = 0T. The sample temperature is ~ 25 mK.

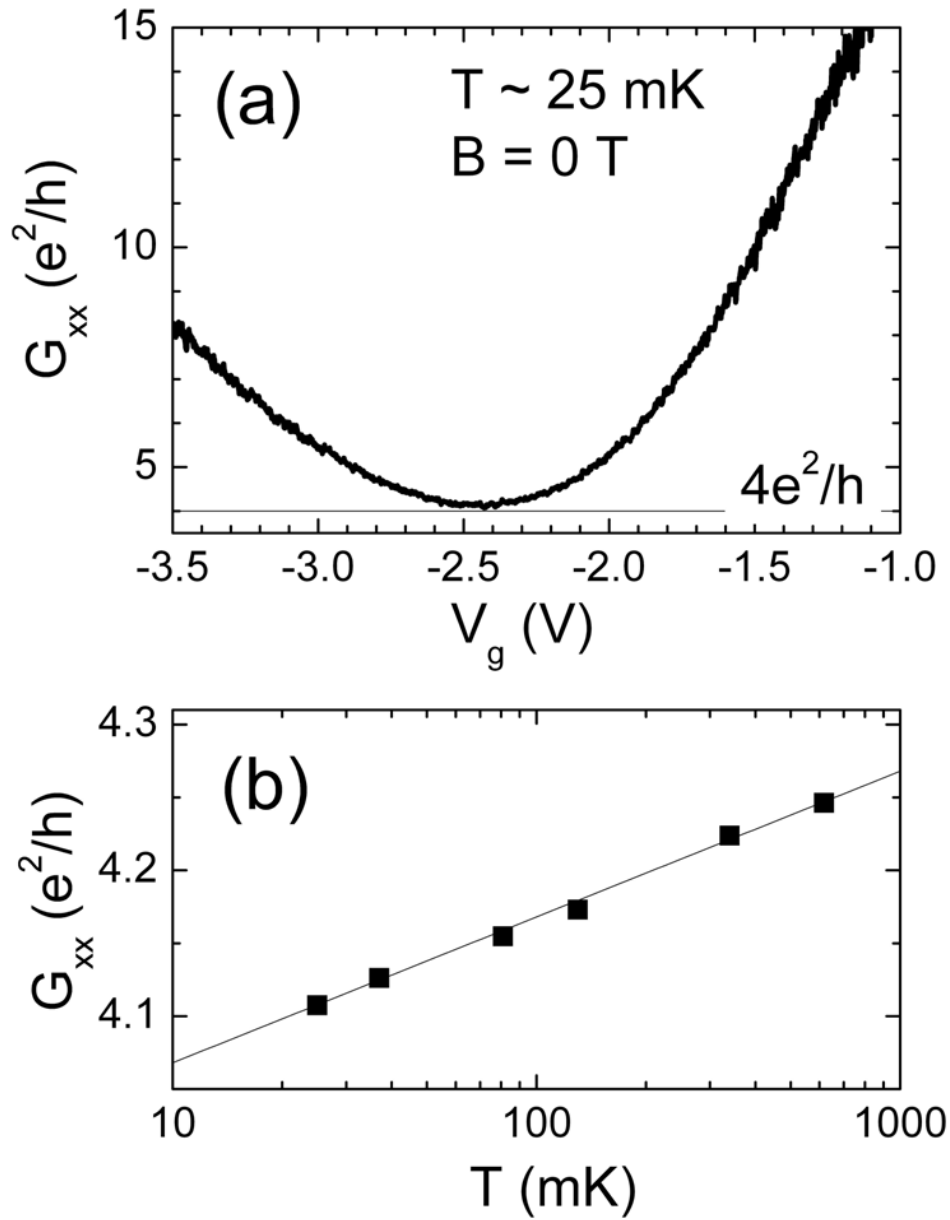

Figure 2 (a) Zero magnetic field conductance $G_{xx}$ near the CNP. The minimal value is close to $4e^2/h$. (b) semi-log plot for the temperature dependence of the minimal conductance. The line is a linear fit.

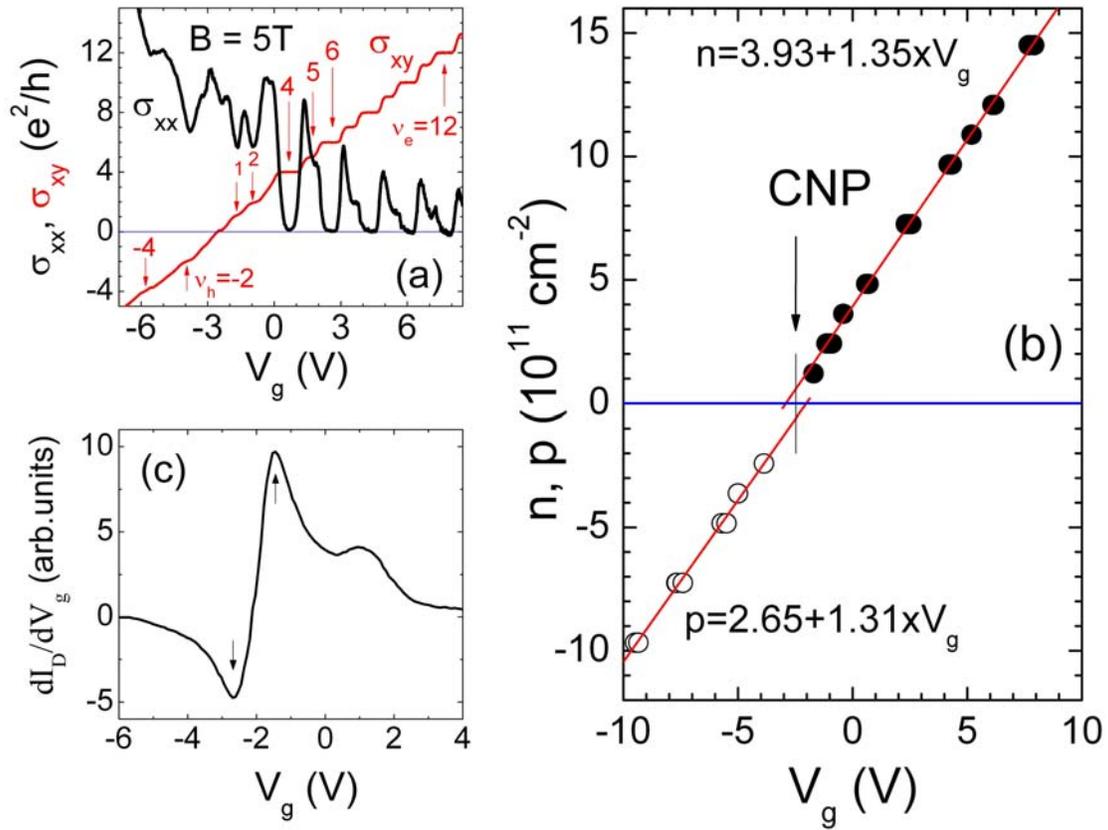

Figure 3 (color online) (a) $\sigma_{xx}$ and $\sigma_{xy}$ versus $V_g$ at B = 5T. The sample temperature is ~ 30 mK. The $\sigma_{xx}$ trace is amplified by a factor of 10 for clarity. Arrows mark the integer quantum Hall stats in the electron and hole regimes. (b) Electron density (n) and hole density (p) as a function of $V_g$. The charge neutrality point (CNP) is marked. Red lines are a linear fit. (c) Transconductance as a function of gate voltage.

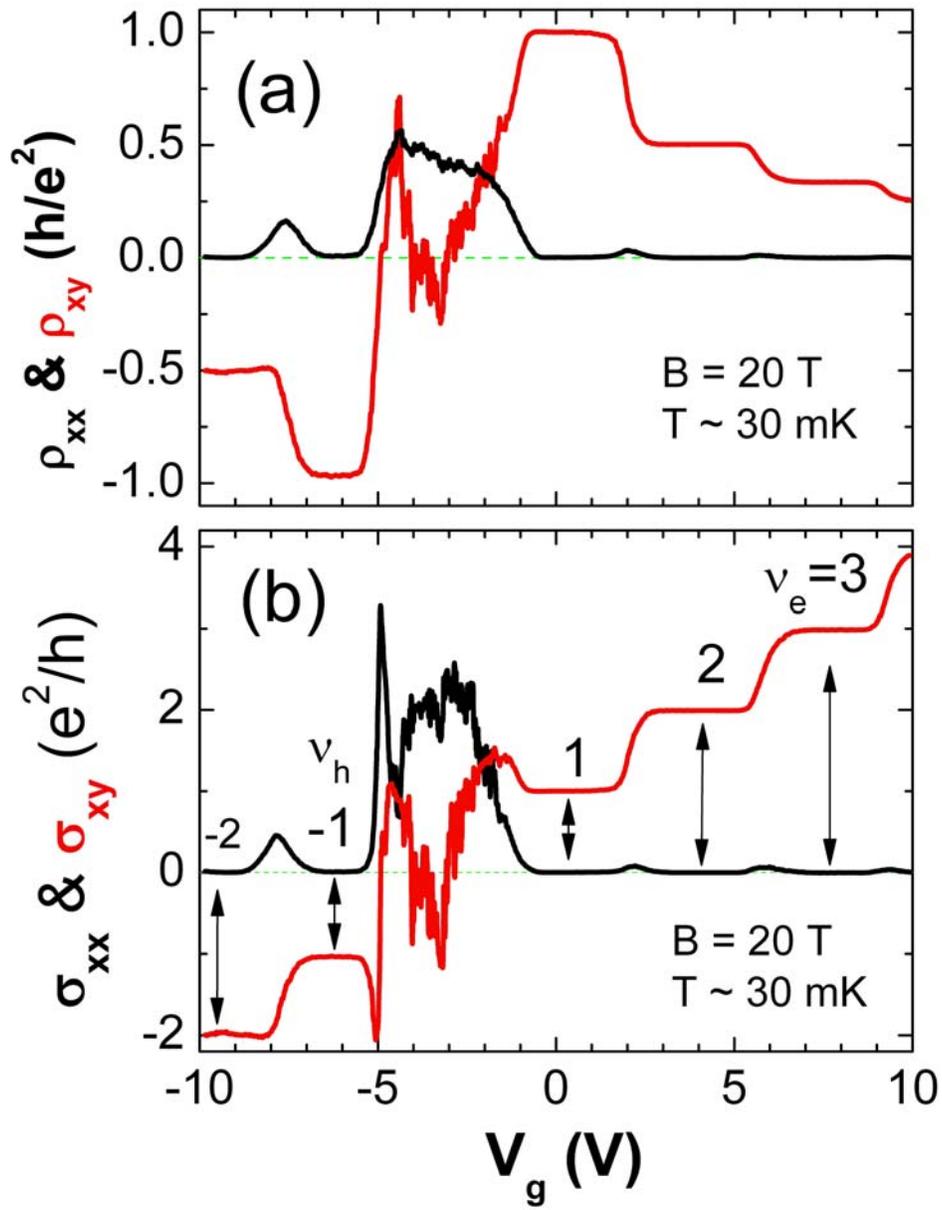

Figure 4 (color online) (a) $\rho_{xx}$ (black trace) and $\rho_{xy}$ (red trace) versus gate voltage at B = 20 T. (b) $\sigma_{xx}$ (black) and $\sigma_{xx}$ (red) versus gate voltage. Arrows mark the positions of the well developed quantum Hall states at $\nu_e$ = 1,2,3 in the electron regime and $\nu_h$ = -1, -2 in the hole regime.

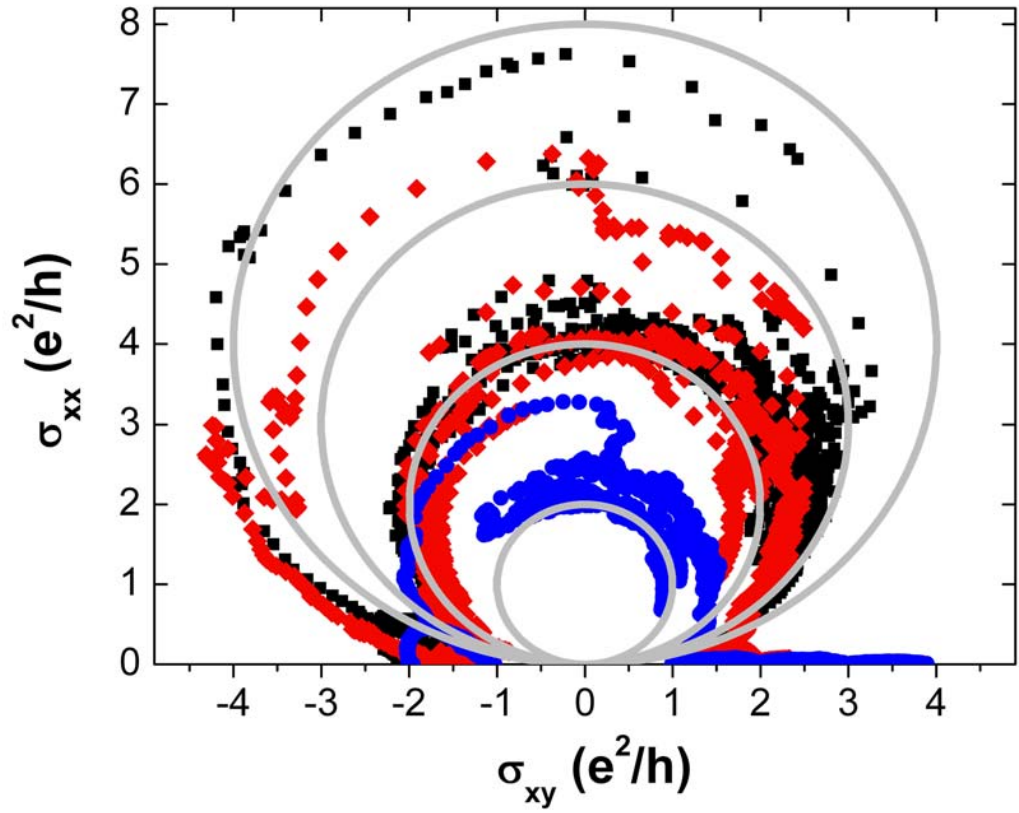

Figure 5 (color online) $\sigma_{xx}$ versus $\sigma_{xy}$ at B = 20T (blue dots), 25T (black squares), and 30T (red diamonds). Gray circles are for $(\sigma_{xx} - N)^2 + \sigma_{xy}^2 = N^2$, with N = 1,2,3,4.